\documentstyle[epsfig,amsmath,amssymb,float]{elsart}

\begin{document}
%\draft
\begin{frontmatter}
\title{Collective dynamics of fermion clouds in cigar-shaped traps}
\author[sns,ist]{Z. Akdeniz}, 
\author[sns]{P. Vignolo} 
and 
\author[sns]{M.P. Tosi\corauthref{cor1}}
\corauth[cor1]{Corresponding author, e-mail: {\tt tosim@sns.it}}
\address[sns]
{NEST-INFM and Classe di Scienze,
Scuola Normale Superiore,\\ I-56126 Pisa, Italy}
\address[ist]
{Department of Physics, University of Istanbul, Istanbul, Turkey}
\maketitle

\begin{abstract}
The propagation of zero sound in a spin-polarized Fermi gas under 
harmonic confinement
is studied as a function of the mean-field
interactions with a second Fermi gas. A local-density
treatment is compared with the numerical solution of the
Vlasov-Landau equations for the propagation of density distortions 
in a trapped
two-component Fermi gas at temperature $T=0.2\,T_F$.
The response of the gas to the sudden creation of a sharp hole at its 
centre is also studied numerically.
\end{abstract}
\begin{keyword} 
Fermi gases \sep  zero sound
\PACS{03.75.Ss\sep05.30.Fk}
\end{keyword}
\end{frontmatter} 

\section{Introduction}
Current experiments on the collisional properties
of a Fermi gas in two spin-polarized states under confinement have proven
that this setup provides an important tool for investigating
the dynamics of dilute quantum gases.
For such a system in harmonic trapping the existence of zero and first sound
excitations has been demonstrated experimentally \cite{gensemer,demarco}
and numerically~\cite{noi}
by measuring the damping of the oscillations
of a non-equilibrium density profile 
as a function of the collision rate.
These studies have followed the evolution of the dynamical behaviour
of the quantum gas from a collisional (first sound) regime to
a collisionless (zero sound) regime. However, no direct measurement
of the speed of sound propagation has been carried out so far in confined
Fermi gases, contrary to the case of Bose-Einstein condensates in
elongated traps~\cite{ketterle} where the propagation of density distortions 
has been followed experimentally.

The main purpose of the present study is to evaluate by both theoretical and
numerical means the propagation of density distortions in the zero-sound
regime for a degenerate Fermi gas contained inside a cigar-shaped
harmonic trap, in situations that are accessible to experiment.
It may be useful to recall
at this point the earlier studies of these collective excitations in 
homogeneous Fermi fluids.
The experiments of Abel {\it et al.}~\cite{abel} on acoustic waves propagating
through liquid $^3$He demonstrated that the transition from collisional 
to collisionless dynamics is signalled by a drastic change in the frequency
dependence of sound-wave attenuation and showed that the difference in 
the speeds of first and zero sound is quite small. This is as expected 
for a strongly coupled Fermi liquid from the Landau theory~\cite{pines}.
The evolution of the zero-sound dispersion relation in liquid $^3$He
into the microscopic regime was determined by Sk\"old {\it et al.}~\cite{skold}
in inelastic neutron scattering experiments and accounted for theoretically
by Aldrich {\it et al.}~\cite{aldrich} within a polarization potential 
approach extending Landau's Fermi-liquid theory to larger momenta and energies.
Landau's theory has also been used by Yip and Ho~\cite{yip-ho} to evaluate the
collisionless excitation modes corresponding to density fluctuations and to 
coherent interconversion of various spin species in a homogeneous Fermi gas.

At variance from the strongly coupled $^3$He liquid,
in a dilute Fermi gas the interactions can be expected to introduce
only slight modifications to the bare sound velocities, which 
in the homogeneous noninteracting limit are 
$v_F$ and $v_F/\sqrt{3}$ for zero and first sound,
respectively. Here, the Fermi velocity $v_F$ is determined
by the Fermi wave number and hence by the particle number density.
These well-known results~\cite{pines} immediately suggest how the 
propagation of collective density fluctuations in a confined Fermi gas
may be treated within a local-density approximation.

In this work we first present an analytical study of the behaviour 
of the Fermi 
velocity and of the speed of zero sound in 
a two-component inhomogeneous Fermi
gas as functions of the mean-field interactions in the collisionless regime.
We support our analytical estimate by numerical simulations
solving the Vlasov-Landau equations (VLE) 
for the fermionic Wigner distributions at 
$T=0.2\,T_F$, $T_F$ being the Fermi temperature~\cite{noi}.
We evaluate the propagation of zero sound by simulating
an experiment where perturbation in the density profile is introduced
by a laser beam permanently focused on the centre of the 
trap~\cite{damski,torma}. 
We also simulate an experiment carried on by Dutton 
{\it et al.}~\cite{hau}
on a Bose-Einstein condensate, where a defect with a large gradient in the
density is generated by
cutting off a slice of the profile by means of an ultra-compressed
slow pulse of light.
In their experiment the generation of quantum shock waves in the
condensate was observed.

The paper is organized as follows. In Sec.~\ref{themodel} we recall the
equations of motion for a two-component Fermi gas and evaluate in a 
semiclassical local-density approach the Fermi velocity and the zero 
sound velocity for such a system. In Sec.~\ref{result} we report 
illustrations of 
our numerical simulations for a one-component and a two-component Fermi gas,
and compare the results with our theoretical estimates.
Conclusions and perspectives are given in Sec.~\ref{concl}.

\section{The model}
\label{themodel}
The two spin-polarized fermionic components of the gas are described 
by the Wigner distribution functions 
$f^{(j)}({\bf r},{\bf p},t)$, with $j=1$ or 2. 
These obey the kinetic equations
\begin{equation}
\dfrac{\partial f^{(j)}}{\partial t}+\dfrac{{\bf p}}{m_j}\cdot{\bf \nabla} 
f^{(j)}-
{\bf \nabla} U^{(j)}\cdot {\bf \nabla}_{{\bf p}}f^{(j)}
=C_{12}[f^{(j)}].
\label{vlasov}
\end{equation}
At high dilution the interactions between atoms of the same spin are suppressed
by the Pauli principle, so that
the mean-field Hartree-Fock (HF) effective potentials are
$U^{(j)}({\bf r},t)= V_{ext}^{(j)}
({\bf r},t)+g n^{(\overline{j})}({\bf r},t)$
with $\overline{j}$ indicating the species different from
$j$ and $V_{ext}^{(j)}({\bf r},t)$ being the sum of external confining
and driving potentials. 
Here we have set $\hbar=1$ and $g=2\pi a/m_r$, with  
$a$ the $s$-wave scattering length ($a>0$ for the present case of 
repulsive interactions) and 
$m_r$ the reduced mass.
The total spatial densities
$n^{(j)}({\bf r},t)=n_0^{(j)}({\bf r})+\delta n^{(j)}({\bf r},t)$ 
are the sum of the equilibrium
densities $n_0^{(j)}({\bf r})$ and the time-dependent perturbations
$\delta n^{(j)}({\bf r},t)$, and are obtained by integration of 
$f^{(j)}({\bf r},{\bf p},t)$ in momentum space.
Finally $C_{12}[f^{(j)}]$ is the collision integral, which will be set
to zero in the collisionless regime. Equation~(\ref{vlasov}) in this
regime will be solved numerically in Sec.~\ref{result}, using
the algorithm developed by Toschi {\it et al.}~\cite{noi} and taking
the equilibrium density profiles as given 
in the HF approach~\cite{madda} by
\begin{equation}
n^{(j)}_{0}({\bf r})=\int\dfrac{d^3p}{(2\pi)^3}
\left\{\exp\left[\beta
\left(\frac{p^2}{2m_j}+U^{(j)}({\bf r})-
\mu^{(j)}\right)\right]+1\right\}^{-1},
\label{equilibrium}
\end{equation}
where $\beta=1/k_BT$ and $\mu^{(j)}$ is
the chemical potential of species $j$.

In the rest of this section we present instead a theoretical
discussion of the long-wavelength excitations of the gas.
The perturbation $\delta n^{(j)}({\bf r},t)$ moves with a group velocity
whose natural unit is the Fermi velocity $v_F^{(j)}$.
In an inhomogeneneous gas the Fermi velocities depend on position and will
be related below to the local chemical potential of each species through
a local density approximation. Assuming for the moment that they are known,
the dynamical
response of the two-component Fermi gas can be calculated
within a random phase approximation~\cite{fetter,yip}
from the set of eigenvalue equations
\begin{equation}
\delta n^{(j)}(q,\Omega)=-\chi_j(q,\Omega)
[\delta V^{(j)}(q,\Omega)+g\delta n^{(\overline{j})}(q,\Omega)]\,,
\label{ito}
\end{equation}
where 
\begin{equation}
\chi_j={\cal N}_j\left[1-\dfrac{\Omega}{2v_F^{(j)}q}\ln
\left(\dfrac{\Omega+v_F^{(j)}q}{\Omega-v_F^{(j)}q}\right)\right]
\label{ato}
\end{equation}
with
${\cal N}_j= v_F^{(j)}m_j^2/2\pi^2$ being the density of states 
of each component per unit volume at the Fermi level.
In writing these equations we have specifically assumed that we are
treating the case of long-wavelength excitations propagating with
wave number $q$ along the axis of an elongated cylindrical trap, with
the Fourier transform $\delta V^{(j)}(q,\Omega)$ of the driving
potentials being independent of the radial coordinate and the Fourier
transform $\delta n^{(j)}(q,\Omega)$ of the density distortions
being taken at the value $r=0$ of the radial coordinate.
Following~\cite{yip-ho} and~\cite{yip} it is easily seen
that the collective modes of the Fermi gas described
by Eq.~(\ref{ito}) are determined by the solutions
of the equation
\begin{equation}
1-g^2\chi_1(q,\Omega)\chi_2(q,\Omega)=0\,\,,
\label{fame}
\end{equation}

yielding in general two eigenfrequencies $\Omega_j(q)$.

Let us turn next to the calculation of the Fermi velocities.
Each of them is determined by the corresponding single-particle 
dispersion relation as $v_F=\partial E(k)/\partial k|_{k=k_F}$, yielding in
a local density approximation
\begin{equation}
v_F^{(j)}=(2\mu^{(j)}_{\rm loc}/m_j)^{1/2}
\end{equation}
where $\mu^{(j)}_{\rm loc}({\bf r})=\mu^{(j)}-U^{(j)}({\bf r})$
is the local chemical potential.
In cylindrical harmonic confinement and in the 
absence of interactions $v_F^{(j)}$ can be written as
\begin{equation}
v_F^{(j)}(r,z)=\left[2\left(\mu^{(j)}-V^{(j)}(r,z)\right)/m_j
\right]^{1/2}
\label{sonno}
\end{equation}
where
$V^{(j)}(r,z)=m_j\omega_j^2(r^2+\varepsilon_j^2z^2)/2$, with 
radial trap frequency $\omega_j$ and anisotropy
$\varepsilon_j$. Under driving the radial and axial components of 
$v_F^{(j)}$ 
then obey harmonic equations of motion,
\begin{equation}
v_F^{(j)}(t)|_{z=0}=\left(2\mu^{(j)}/m_j\right)^{1/2}
\sin(\omega_jt+\phi_r)
\end{equation}
and
\begin{equation}
v_F^{(j)}(t)|_{r=0}=\left(2\mu^{(j)}/m_j\right)^{1/2}
\sin(\varepsilon_j\omega_jt+\phi_z),
\label{holiday}
\end{equation}
where the phases $\phi_r$ and $\phi_z$ are determined by
the initial value of
the perturbation.
In the presence of mean-field interactions, on the other hand, 
Eq.~(\ref{sonno}) is replaced by
\begin{equation}
v_F^{(j)}(r,z)=\left[2\left(\mu^{(j)}-V^{(j)}(r,z)
-gn_0^{(\overline j)}(r,z)\right)/m_j
\right]^{1/2}.
\label{bige}
\end{equation}
In this case
harmonicity under driving will be
lost, except when
the mean field term can be treated perturbatively
and for small axial 
displacements from to the center of the trap.
In this regime it is easy to demonstrate that the mean-field interactions 
lower the Fermi velocity. More generally, their net effect is to decrease 
the chemical potential by inducing a decrease of the fermion density.

With the above as background, we conclude this section by presenting
our estimates for the speed $c_0$ of collisionless sound waves propagating
along the trap axis at $r=0$ in two special cases, that will be tested in
the next section by comparison with the evolution of density profiles
in time from the numerical solution of Eq.~(\ref{vlasov}):

(i) {\it A single-component Fermi gas.} Setting $g=0$, Eqs.~(\ref{ito})
and~(\ref{ato}) immediately yield
\begin{equation} 
c_o=v_F(0,z).
\end{equation}
This situation will arise not only in the obvious case where the
interspecies mean-field interactions vanish, but also in the case where they
become so strong in a two-component Fermi gas that the regime of phase 
separation predicted by the theory~\cite{madda} is being approached.
In this latter case the residual mean-field interactions with the second 
species merely shift the chemical potential of the species that resides in 
the inner part of the trap and hence will shift its Fermi velocity.

(ii) {\it A symmetric two-component Fermi gas.}
Taking equal masses and numbers of particles for the two fermionic species 
inside identical traps, we can write $c_0=\lambda v_F$ for the local speed 
of sound wave propagation and determine $\lambda$ from Eq.~(\ref{fame})
using the Landau Fermi-liquid theory to describe the mean-field interactions
through the appropriate Landau parameters.
Only the spin up-down scattering is allowed in the {\it s}-wave channel for
a two-component mixture of spin polarized fermions, so that the only 
non-zero interaction parameters are $F_0^s$ and $F_0^a$ with the
condition $F^s_0=-F^a_0$ ($=F_0$, say). Equation~(\ref{fame})
can be written in the form
\begin{equation}
\dfrac{\lambda}{2}
\ln\left(\dfrac{\lambda+1}{\lambda-1}\right)=1+\dfrac{1}{F_0}
\label{phase-sep}
\end{equation}  
with 
\begin{equation}
F_0=g{\cal N}=2a k_F/\pi.
\label{weekend} 
\end{equation}

\begin{figure}[H]
\centering{
\epsfig{file=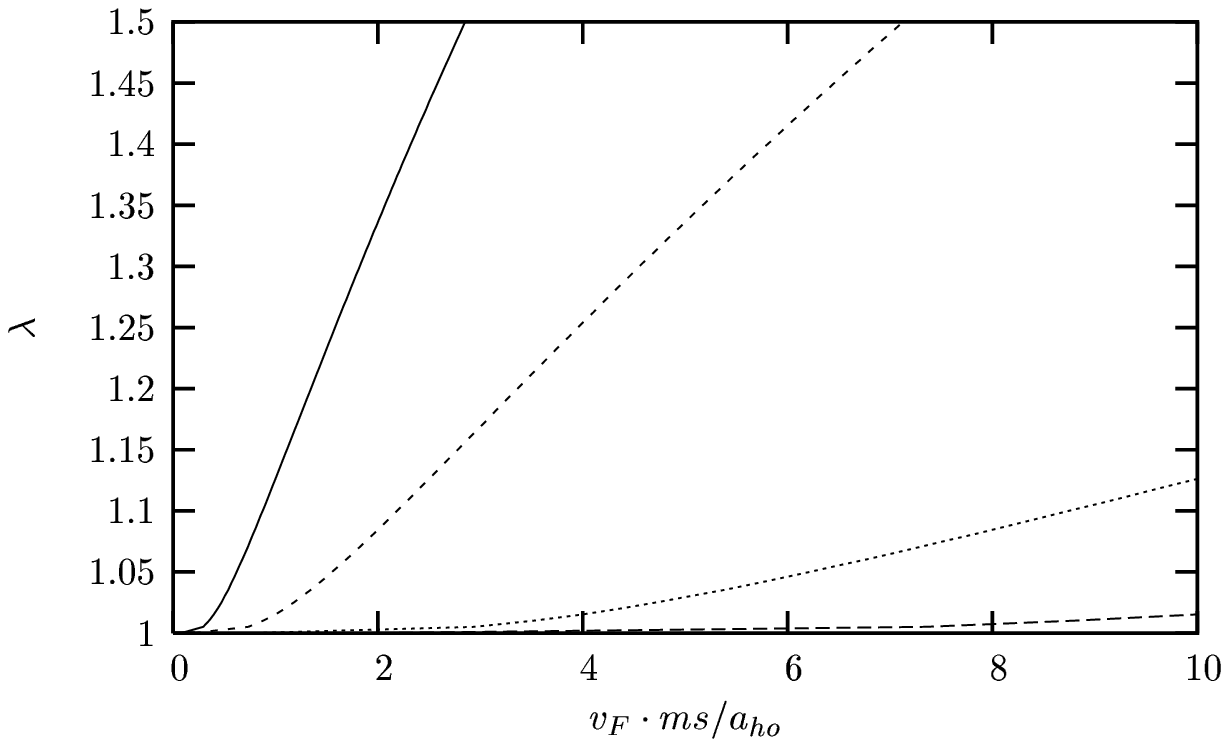,width=1\linewidth}}
\caption{The $\lambda$ factor as a function of $v_F$ (in units of $a_{ho}/ms$
with $a_{ho}=(\hbar/m\omega)^{1/2}$) for various values of the scattering 
length: $a=2\times 10^4\,a_0$ (continuous line), 
$8\times 10^3\,a_0$ (short-dashed line), 
$2\times 10^3\,a_0$ (dotted line) and
$800\,a_0$  (long-dashed line), $a_0$ being the Bohr radius.}
\label{fig0}
\end{figure}

In Figure~\ref{fig0} we have plotted the behaviour of the solution
of Eq.~(\ref{phase-sep}) as a function of the Fermi velocity for
various values of the scattering length.
We also show in Figure~\ref{fig0_1} the values of $v_F$ and $c_0$
as functions of the scattering length, for the choice
of the other system parameters that will be adopted in the numerical
calculations reported in the next section.
It is seen from these figures
that, in order to obtain a very significant increase of the speed of zero
sound over the Fermi velocity from repulsive mean-field interactions, one would
need to work on a high-density system having huge values of the scattering 
length.
The value of $a$ can to some extent be tuned by exploiting
Feshbach resonances~\cite{fesh}, but strong repulsions will also
favour demixing of the two-component Fermi gas.

\section{Numerical results}
\label{result}
\subsection{Perturbation introduced by a permanent laser beam}
As a first case we consider a one-component 
gas of 1000 spin-polarized $^{40}$K atoms inside
an axially
symmetric trap with $\omega=100$\,\,s$^{-1}$ and $\varepsilon=0.1$.
We use 200 computational particles 
to represent each fermion, in order to obtain a 
low noise-to-signal ratio~\cite{noi}.
The equilibrium density profile is perturbed by 
an effective potential 
$U_0\exp(-z^2/2\sigma^2)$ with amplitude $U_0$ and width $\sigma$,
which simulates a laser beam acting continuously on the centre of the trap. 
The fermions interacting with the laser are pushed out of 
the trap centre and the density distortion moves towards the ends of the trap
with a velocity which corresponds
to the local Fermi velocity given by Eq. (\ref{sonno}).
Figure~\ref{fig1} shows the evolution in time of the perturbation
along the $z$ axis for $r=0$. For such a system the maximum value
of the axial zero-sound velocity, which is equal
to the axial Fermi velocity, can be evaluated using
Eq.~(\ref{sonno}) as $c_0^{max}|_{r=0}=v_F^{max}|_{r=0}=0.383\,a_{ho}/ms$,
where $a_{ho}=(\hbar/m\omega)^{1/2}$ is the radial harmonic oscillator
length.
Figure~\ref{fig1} also shows that the positions of the density peaks 
as calculated numerically at various time steps
are in good agreement with those estimated from Eq.~(\ref{sonno}). 

\begin{figure}[H]
\centering{
\epsfig{file=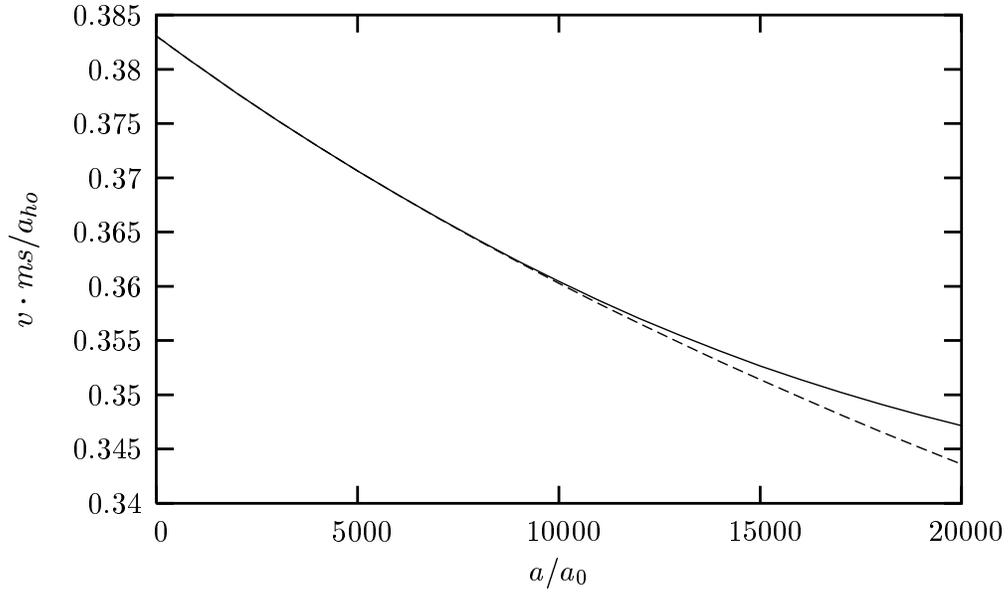,width=1\linewidth}}
\caption{The speed of zero sound (continuous line) and the Fermi 
velocity (dashed line), in units of $a_{ho}/ms$, as functions of the 
scattering length
$a$ (in units of $a_0$)
for a two-component Fermi gas. 
The calculation refers to 2000 particles 
in an anisotropic harmonic trap with $\omega=100$\, s$^{-1}$
and $\varepsilon=0.1$ and the values of the velocities
refer to the case of a little perturbation near the centre of the trap.}
\label{fig0_1}
\end{figure}

\begin{figure}[H]
\centering{
\epsfig{file=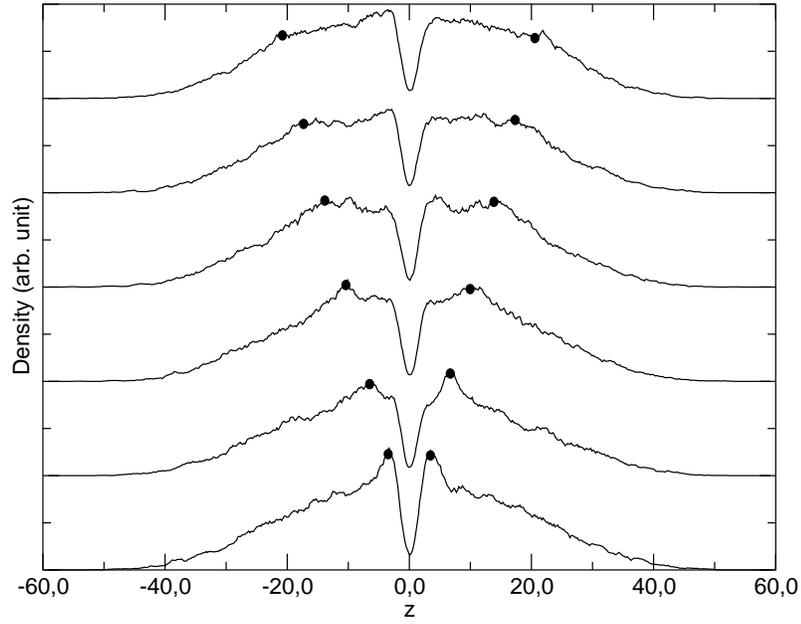,height=1\linewidth,angle=270}}
\caption{Time evolution of a defect introduced in a one-component 
Fermi gas, plotted as a function of the axial distance $z$ 
(in units of $a_{ho}$). The bottom profile refers to the initial
state and the subsequent profiles are separated by time intervals
of $10\,ms$.
The dots indicate the analytical estimate 
of the position of 
the perturbation peak from Eq.~(\ref{sonno}).}
\label{fig1}
\end{figure}

\begin{figure}
\centering{
\epsfig{file=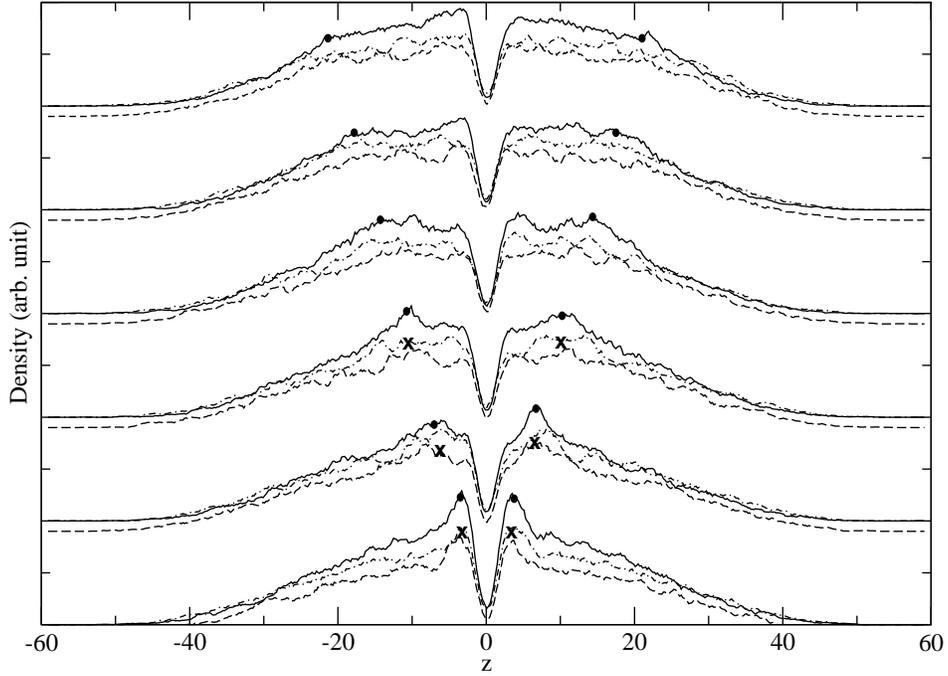,height=1.1\linewidth,angle=270}}
\caption{Time evolution of a defect introduced in both components of an
interacting two-component Fermi gas
with $a=2\times 10^4\, a_0$ (dashed lines),
in comparison with the one-component case (continuous lines).
From bottom to top each profile corresponds to a time lag of 10 ms.
Dots indicate the analytical estimate 
of the position of the peak in the non-interacting case from  
Eq.~(\ref{sonno}) and crosses
indicate the peak positions corresponding to the zero sound velocity
evaluated at $t=0$ from 
Eqs.~(\ref{bige}), (\ref{phase-sep}) and (\ref{weekend}).}
\label{fig2}
\end{figure}

We consider next a 50-50 mixture of 2000
$^{40}$K atoms which are polarized in the $7/2$ and $9/2$ spin states,
under the same external potentials as in the previous case.
The value of the scattering length of $^{40}$K ($a=80$ Bohr radii $\,a_0$)
is so small that $v_F$ is lowered by the interactions by only 
0.1$\%$. From Fig.~\ref{fig0_1} it is evident that the speed of zero sound is practically unchanged.
To see the interaction effects we study the case $a=2\times 10^4\,a_0$.
For this value of the scattering length 
the Fermi velocity is obtained from the numerical
calculation of the chemical potential and of the effective potential
as $v_F^{max}|_{r=0}=0.343\,a_{ho}/ms$.
The corresponding values of $\lambda$ and $c_0$ are  1.01 
and $0.347\,a_{ho}/ms$ respectively (see Fig.~\ref{fig0_1}).
The time evolution of the perturbation in the two components of
the interacting Fermi gas is shown in Fig.~\ref{fig2}, in comparison with
that of the noninteracting gas.

\begin{figure}
\centering{
\epsfig{file=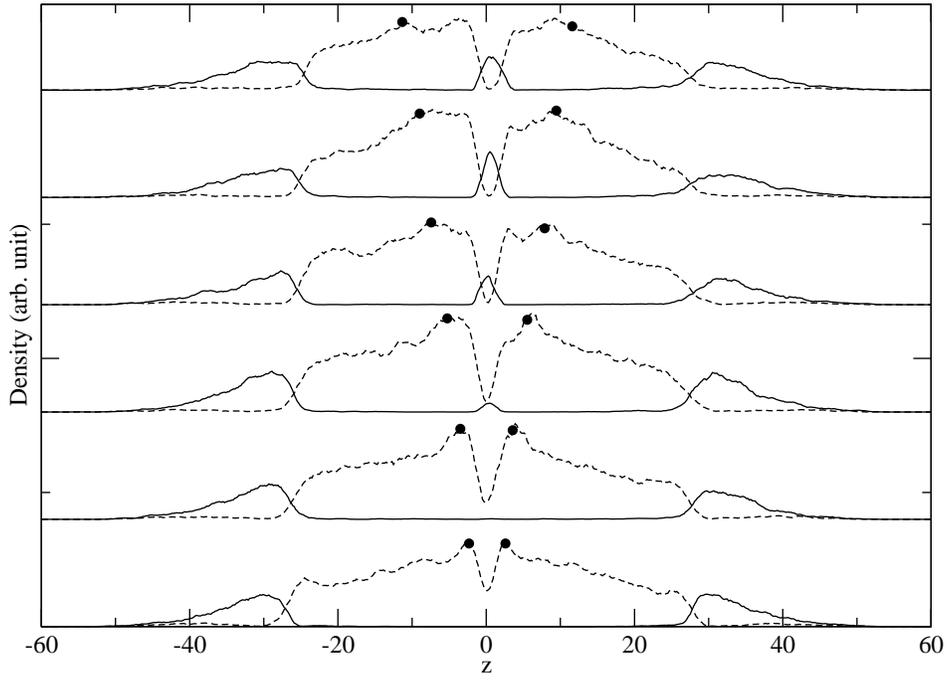,height=1.1\linewidth,angle=270}}
\caption{Time evolution of a defect introduced in one component 
of an
interacting two-component Fermi gas 
with $a=8\times 10^4\, a_0$ in a
regime of phase separation (dashed lines).
From bottom to top each profile corresponds to a time lag of 5 ms.
The dots indicate the analytical estimate 
of the position
of the peak of the perturbation from Eq.~(\ref{sonno}).}
\label{fig3}
\end{figure}

At still higher couplings spatial phase separation starts in the
two-component Fermi gas. The case $a=8\times 10^4\, a_0$ is 
illustrated in
Figure~\ref{fig3}. The two components are amost completely demixed
and only one of them feels the external perturbation. In this case
the interactions with the second component merely shift the chemical
potential entering the evaluation of the Fermi velocity of the first
component, which equals the speed of propagation of zero sound.

\subsection{Perturbation introduced by a compressed slow light pulse}
In the work of Dutton {\it et al.}~\cite{hau} 
ultra-compressed slow light-pulses
were used
to pump a small slice of a Bose-Einstein condensate into an untrapped state,
leaving behind a sharp hole in the density profile.
The density defect in the condensate evolves
through the formation of solitons,
this process being analogous to the creation of
shock waves in a classical fluid.

\begin{figure}
\centering{
\epsfig{file=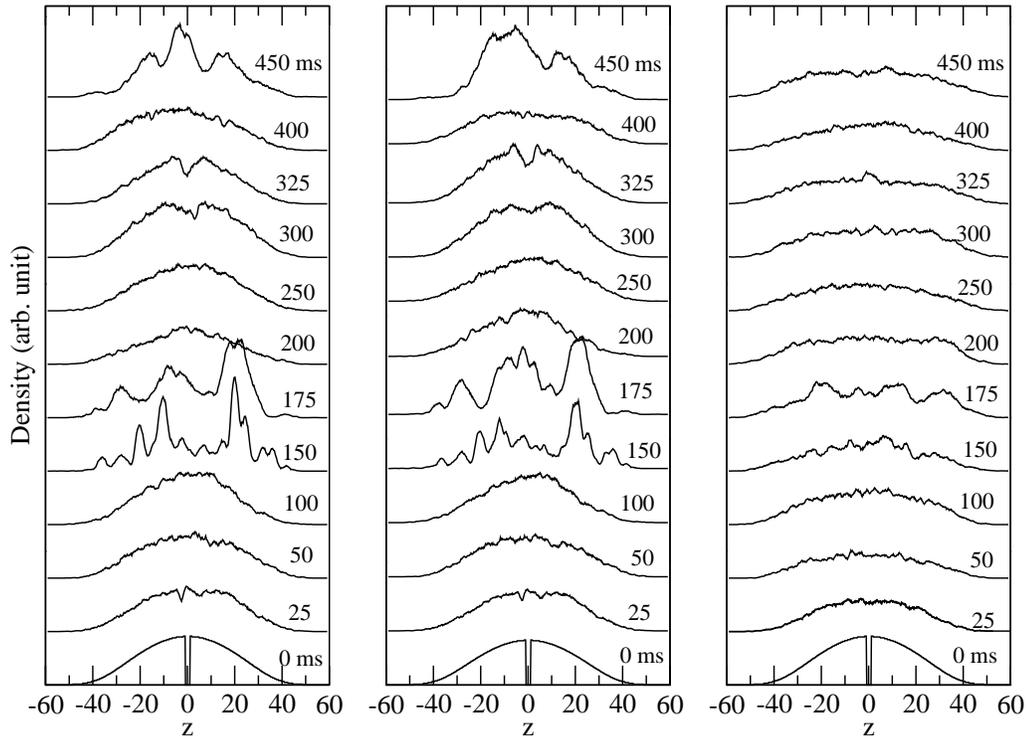,height=1.2\linewidth,angle=270}}
\caption{Time evolution of the axial density profile as a function
of $z$ (in units of $a_{ho}$) for the case of a small U-shape vacancy
with $a=0$, $800\,a_0$ and $2\times10^4\,a_0$ (from left to right).
From bottom to top, the time lag of each profile (in ms)
is indicated in the figure. The profiles which are not shown in 
the ranges of time lag
between 200 and 300\,ms and between 325 and 450\, ms are essentially smooth.}
\label{fig5}
\end{figure}

We generate the same type of initial profile in a two-component Fermi gas
for various shapes of the density vacancy (V-shape or 
U-shape of various sizes) and examine the effects of the density 
gradient and of the
size of the defect.
No coherent behaviour is observed in the time evolution
of the density profiles.
In Fig.~\ref{fig5} we plot the time evolution of one
of the two components of the Fermi gas for various values
of the scattering length ($a=0$,
$800\,a_0$ and $2\times10^4a_0$) in the case of a small U-shape vacancy.
In the two former cases 
the particles in filling the vacancy generate some density waves,
which reach a maximum amplitude after about 150 $ms$.
This is the time lag nedeed 
for the perturbation to arrive at the ends of the trap and to be 
reflected there, as 
can be roughly estimated by dividing the axial half-length $L_z$
of the trap by $c_0^{max}$ giving $L_z/c_0^{max}
\simeq 140$\,ms.
Because of the absence of thermal dissipation in our model,
the density waves continue to travel along the profile and are
again reflected at the ends of the trap with a periodicity
of about $2L_z/c_0^{max}\simeq 280$\,ms.
The revivals of smaller amplitude which are seen to occur at 
about 300\,ms
correspond to the situation in which the perturbation is 
passing through the centre of the trap.
On increasing the coupling strength up to $a=2\times 10^4\,a_0$
(right panel in Fig.~\ref{fig5}), we find that
the restoring force due to the other fermionic component
plays a role in depressing the intensity of the density waves
relative to those seen in the weak-coupling regime.

\section{Conclusions}
\label{concl}

In summary, we have studied the propagation of collisionless sound
waves in a trapped two-component Fermi gas, assuming a mean-field
model for the repulsive interactions between the two fermionic
species and neglecting dissipative collisions as described by a
Boltzmann quantum collision integral. The inclusion of such collisions
is needed to examine the transition from the zero-sound to
the first-sound regime~\cite{noi}, that we expect will be characterized 
by a marked
change in the speed of sound propagation. We also expect that zero-sound
waves will be Landau-damped in the case of attractive interactions,
as is the case for a homogeneous Fermi liquid~\cite{pines}.

We have found that, both at weak coupling and in the strong-coupling
regime where the two Fermi clouds are demixing from their mutual
repulsions, the speed of zero-sound waves is very close to the 
Fermi velocity as determined by the local particle density through
the local chemical potential. These results have been obtained
from comparing semi-analytical estimates with numerical studies
of the evolution of peaks in the density profiles under driving
from an external potential simulating a blue-detuned laser that
keeps expelling atoms from the central region of the trap.

We have also examined numerically the time evolution of impulsively
created density defects which are characterized by sharp density
gradients in a two-component Fermi gas with repulsive interactions.
In the collisionless regime we have found that the filling
of the vacancy at the trap centre generates density waves which
travel to the ends of the trap and are back-reflected there, so
that they appear as periodic revivals of distortions in
the density profile. These revivals become attenuated as the strength
of the coupling is increased, presumably as a consequence of the 
increasing efficiency of momentum-redistribution processes through
non-dissipative collisions at strong coupling.

\ack
We acknowledge support from INFM through PRA2001-Photonmatter.
ZA also acknowledges support from TUBITAK and from the Research Fund of the
University of Istanbul under Project Number BYP-110/12122002.

\newpage
\newpage
\newpage

\end{document}